**Title:**

# Do we practice what we preach? The dissonance between resilience understanding and measurement

**Authors:** Lukas Halekotte[+°], Andrea Mentges[+*°], Daniel Lichte[+]

[+] German Aerospace Center (DLR), Institute for the Protection of Terrestrial Infrastructures, Rathausallee 12, 53757 Sankt Augustin, Germany

* Corresponding author
*E-mail address:* andrea.mentges@dlr.de

° Co-first authors

**Abstract:**
Resilience is needed to make infrastructures fit for the future, but its operationalization is still lively discussed. Here, we identify three understandings of resilience from the existing literature: resilience as a process, an outcome, and a capacity. We show that all three understandings have their justification, as each plays its part in the core business of resilience, that is, dealing with disruptive events. But, we also find that the trio differs considerably in terms of the implications for the operationalization of resilience. Most importantly, only the understanding of resilience as a capacity allows for a continuous resilience monitoring and a management which is agnostic to the type of disruptive event. We therefore advocate to understand and assess resilience as a capacity. While this understanding is in line with popular opinion, it is often not reflected in the assessment approaches applied. This dissonance shows, for example, in the use of single performance curves to assess resilience. We argue that in order to assess resilience as a capacity, we need to consider multiple performance curves, otherwise we will capture the system's ability to deal with one specific event instead of its ability to deal with any surprises that come its way.

**Keywords:**
Resilience definition; Resilience capacities; Resilience management; Resilience assessment; Performance curves; Critical infrastructures

# 1. Introduction

Today, human-made and natural systems face a diverse suite of sometimes unpredictable or unexpected adverse events, for example, as consequences of climate change, increasing interdependencies, or intentional acts (Middleton and Latty 2016; Meerow et al. 2016; Wieland et al. 2023). As many of these events are unavoidable, coping with them efficiently – i.e., being resilient or acting resilient – is key. Especially for critical infrastructures, the backbone of society (Nick et al. 2023), it is vital to continue delivering essential services to society, no matter what. To achieve this, resilience is often mentioned as the aspired goal that should allow infrastructures to deal with surprises and increase security of supply in times of change (United Nations 2015; European Union 2022).

To date, there is a lot of variability, inconsistency, and disagreement around the term resilience (Jore 2020; Mottahedi et al. 2021; Mentges et al. 2023) and a rich body of literature exists which discusses differences among resilience definitions (Wied et al. 2020; Mottahedi et al. 2021; Nipa et al. 2023; Meerow et al. 2016), the scope of resilience (i.e., what aspects are included in resilience) (Jackson and Ferris 2013), the history of the term resilience (Alexander 2013), or the various existing approaches to measure resilience (Bi et al. 2023). What is largely missing, however, is a thorough discussion of the understanding of resilience on a much more general level: Do we think that resilience is an ability a system can acquire, a process a system can conduct or a measure of a system's performance under pressure? And what does this imply for the assessment and management of resilience?

We suggest that there are three main resilience understandings. First, resilience is often understood as a capacity (Carlson et al. 2012; Bruneau et al. 2003; NIAC 2010; Rose 2007; Ouyang and Wang 2015; Cai et al. 2018; Béné et al. 2012; Walker et al. 2004; Fiksel 2006; UNISDR 2015). In this case, resilience is introduced as a skillset or a suite of abilities which allows the system to deal with disruptions efficiently. Second, resilience can also be understood as the reactive process which follows the disruption of a system (Jackson and Ferris 2013; Kanno et al. 2019; Norris et al. 2008; Ungar 2018; Scholl and Patin 2014). In this understanding, resilience would describe the series of actions that are performed to counteract a disruptive event. Third, resilience can be seen as an outcome or measure of the outcome, e.g., a speed, rate, or degree (Diao et al. 2016; Pagano et al. 2019; Tilman and Downing 1994; Alberti et al. 2003; Pimm 1984). This means that resilience would correspond to the measurable "end results" (Gilbert 2010) of the interplay between a disruptive event and the actions taken to counteract it. Finally, some authors understand resilience as a mixture of these main understandings, e.g., they state that resilience is both, a "power or ability" (i.e., capacity) and an "action or act" (i.e., process) (Kanno et al. 2019), or they separately describe resilience as both an "ability" and a "measure" of an outcome (Kamissoko et al. 2023).

A previous approach to distinguish interpretations of resilience groups works which focus on (1) the characteristics which make a system resilient, (2) the actions a system needs to carry out to behave resilient, or (3) the desirable outcome(s) a system should achieve (Moser et al. 2019, see supplement section S1 for a detailed comparison of their and our classification). Here, we go one step further and distinguish between the claimed resilience understanding (reflected in the resilience definition) and the actually applied approach to measure resilience (reflected in the quantification method). We argue that, first, the applied assessment approach should match the claimed understanding, and, second, that the most suitable understanding of resilience is the one which best serves the intended management goal (i.e., managing resilience should lead to infrastructure systems which better cope with any disruptive events). In this context, we demonstrate that the understanding of resilience not only influences the way in which resilience can be measured but also at what points in time and with which resolution such measurements can be conducted. Accordingly, a different understanding also results in different options for the management of resilience.

In the remainder of this work, we first focus on the causal relationship between system/resilience properties, capacities and processes, and on how they affect the outcome of a disruptive event – what we call the "resilience backbone" (section 2). Based on this, we illustrate how some of the most common assessment approaches obtain an estimate of resilience and explain what understanding of resilience they actually address (section 3). We discuss the differences between the understandings, and explain their restrictions and benefits for assessment and management in the context of critical infrastructure research (section 4). Importantly, we see a dissonance between the most common resilience assessment approaches ("what we practice") and the claimed understanding of resilience ("what we preach") – i.e., sometimes, the stated resilience understanding is not really reflected in the resilience measurement (section 5). We further explain why we think resilience is best understood as a capacity, although this poses major challenges for assessing it (e.g., an assessment is only possible via inference). Finally, we outline how the understanding of resilience as a capacity helps to assess and manage resilience.

## 2. Background: The resilience backbone

In order to better understand the relationship between the different understandings, we first introduce our "resilience backbone" as a working basis for the further considerations in this work. The resilience backbone connects all three understandings within a conceptual framework that shows how each of them takes part in the core business of resilience: Coping with disruptive events.

The fundamental components of the resilience concept are a system's properties, capacities, processes and performance. This "resilience backbone" can be characterized by the following widely accepted definitions: A ***system*** is a set of components which act together in order to perform a specific function or a number of specific functions (Mentges et al. 2023). A system is characterized by certain (system) ***properties*** or qualities. The interplay of all its properties determine what the system is able to do, i.e., what ***capacities*** or abilities the system has. What the system actually does – i.e., the ***processes*** or actions the system conducts – is the realization of these capacities. Ultimately, how well the system fulfills its function(s) – i.e., the system's ***performance*** – is an outcome of the processes, given the environment in which the system is situated (see, e.g., Tamberg et al. 2022). An important difference between properties, capacities and processes is how and when they can be observed. System properties constitute the current state or configuration of a system and are therefore observable at any time. System processes, on the other hand, can only be observed when the corresponding actions are carried out (i.e., when the process takes place). System capacities cannot be observed directly, instead conclusions about them can be drawn from observed processes (i.e., from their manifestations).

These agreed-upon, general terms can easily be transferred to their resilience counterparts (e.g., system processes become resilience processes) by relating them to the adverse influences which the system (potentially) has to deal with – i.e., with regard to these influences, the system should be or act resilient. Generally, ***disruptive events*** constitute unfavorable environmental conditions to a system which threaten the process by which it fulfills its function, i.e., they can cause unfavorable or undesired outcomes such as a loss of performance. The actions which a system conducts (i.e., its behavior) to counter a particular disruptive event are the ***resilience processes***. It is these processes which determine how well a system handles a particular disruptive event. Accordingly, the ***outcome*** of a specific disruptive event depends on the interplay between the event and the initiated resilience processes. In this regard, the resilience processes represent the event-specific manifestation of the system's (theoretic) abilities, skills or response options which it can utilize to deal with adverse influences. We refer to these abilities as a system's ***resilience capacities***. This set of skills is again a result of the combination of properties of the system – we refer to the properties which affect a system's resilience capacities as its resilience-building system properties or ***resilience properties***. In contrast to the more abstract resilience capacities, the resilience properties describe tangible features of the system (i.e., they refer to the system configuration). The distinctive feature of resilience properties, resilience capacities and resilience processes, compared to general system properties, capacities and processes, is the direct reference to disruptive events or adverse influences – i.e., they characterize a system which is out of its regular conditions. This direct reference has implications for their observability. While resilience properties can still be observed at any time, resilience processes only occur in response to an adverse event.

The components of the "resilience backbone" are founded on previous research: all terms have previously been used to refer to certain aspects of resilience – see, e.g., Gasser et al. (2019), Wang et al. (2024), or Levine (2014) for "resilience processes", Moser et al. (2019), Gilbert (2010), or references in section 3.1 for "outcome", and Trucco and Petrenj (2023), or references in section 3.2 for "resilience properties". First steps towards describing how these ingredients relate to each other have been taken (see Panteli et al. 2017; Gerges et al. 2023; Moser et al. 2019; Wied et al. 2020). However, a joint conceptual framework – like our resilience backbone – that combines all of these aspects has not yet been presented. Basically, the resilience backbone shows that all three resilience understandings have their justification: each plays a role in the causal chain that describes how a system is capable of coping with a particular disruptive event (see Box 1). We use this framework and its associated terminology to outline the implications of understanding resilience as a capacity, a process or an outcome.

**Box 1: Illustrating the role of resilience capacities and properties in the reaction to a disruptive event**

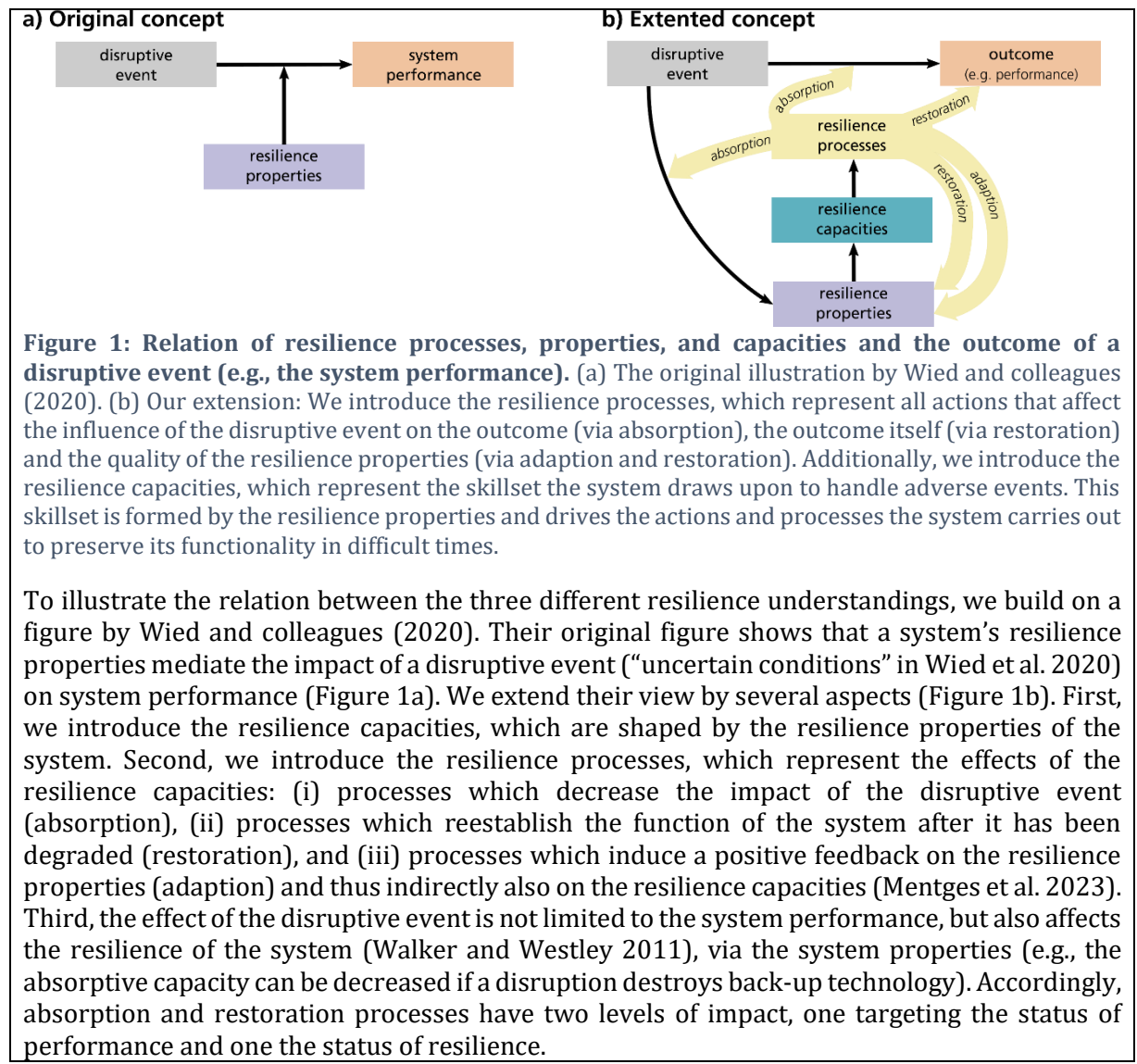

**Figure 1: Relation of resilience processes, properties, and capacities and the outcome of a disruptive event (e.g., the system performance).** (a) The original illustration by Wied and colleagues (2020). (b) Our extension: We introduce the resilience processes, which represent all actions that affect the influence of the disruptive event on the outcome (via absorption), the outcome itself (via restoration) and the quality of the resilience properties (via adaption and restoration). Additionally, we introduce the resilience capacities, which represent the skillset the system draws upon to handle adverse events. This skillset is formed by the resilience properties and drives the actions and processes the system carries out to preserve its functionality in difficult times.

To illustrate the relation between the three different resilience understandings, we build on a figure by Wied and colleagues (2020). Their original figure shows that a system's resilience properties mediate the impact of a disruptive event ("uncertain conditions" in Wied et al. 2020) on system performance (Figure 1a). We extend their view by several aspects (Figure 1b). First, we introduce the resilience capacities, which are shaped by the resilience properties of the system. Second, we introduce the resilience processes, which represent the effects of the resilience capacities: (i) processes which decrease the impact of the disruptive event (absorption), (ii) processes which reestablish the function of the system after it has been degraded (restoration), and (iii) processes which induce a positive feedback on the resilience properties (adaption) and thus indirectly also on the resilience capacities (Mentges et al. 2023). Third, the effect of the disruptive event is not limited to the system performance, but also affects the resilience of the system (Walker and Westley 2011), via the system properties (e.g., the absorptive capacity can be decreased if a disruption destroys back-up technology). Accordingly, absorption and restoration processes have two levels of impact, one targeting the status of performance and one the status of resilience.

# 3. What is practiced? Resilience assessment approaches

Resilience assessment approaches can be categorized based on various criteria (Leštáková et al. 2024), e.g., based on the applied measurement technique (Cutter 2016), based on the theoretical perspective they follow (Rus et al. 2018), based on the dimension of resilience (e.g., social, economic, institutional) they consider (Saja et al. 2019), or based on the capacities (Asadzadeh et al. 2017) or functions (Leštáková et al. 2024) they focus on. One of the most general distinctions can be drawn between outcome (performance-based) and property-based (attribute-based) approaches (Mugume et al. 2015; Diao et al. 2016; Pagano et al. 2019; Förster et al. 2019; Wang et al. 2023). The two approaches differ with regard to 'what' is measured and with regard to 'when' the corresponding measurement can be elicited (Cariolet et al. 2019). The outcome-based approach refers to measurements which quantify how well a system has responded and dealt with one particular or multiple adverse events based on the severity of the event's impact (the 'what'), i.e., this approach addresses resilience "bottom-up" (Figure 2). In order to be applied, it thus generally relies on the occurrence of (an) adverse event(s) (the 'when'). The property-based approach refers to measurements which consider the presence and/or quantify the expression of certain system properties which are assumed to build the system's potential for dealing with adverse influences in a preferable manner (the 'what'), i.e., this approach addresses resilience "top-down" (Figure 2). It can thus potentially be conducted at any time, although it is usually carried out during phases where the system is unaffected by any event (the 'when').

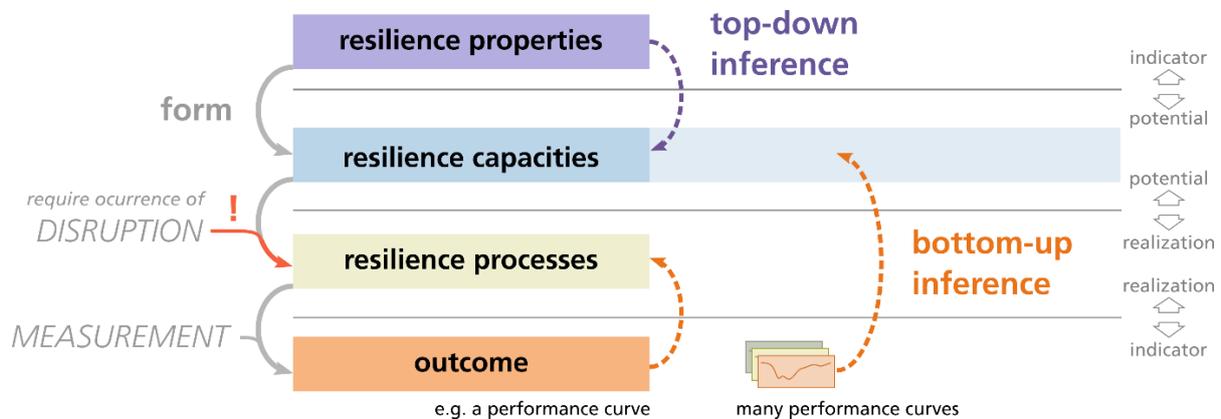

**Figure 2: The relation between property- and outcome-based assessments.** The resilience properties of the system form the top of the cause-effect-chain. They thus affect all aspects below. Therefore, the resilience properties can serve as indicators for the resilience capacities. While the resilience capacities are always present, the resilience processes only occur under the condition that a disruption happened. In the case of a disruption, the effects of the resilience processes may be observed in the performance curve, i.e., the performance curve can serve as an indicator for the resilience processes. Depending on which resilience aspect is estimated, the data required for inference comprise either a single performance curve (to estimate resilience processes) or multiple curves (to estimate resilience capacities).

## 3.1 The outcome-based approach

To present the outcome of a disruptive event, performance curves are certainly the most popular approach in the area of infrastructure protection (see, e.g., Bruneau et al. 2003; Nan and Sansavini 2017; Panteli et al. 2017; Monsalve and La Llera 2019; Argyroudis et al. 2020; Singh et al. 2022; Mentges et al. 2023 , although alternative approaches exist, Rathnayaka et al. 2024; AminShokravi and Heravi 2024; Ganin et al. 2017). A performance curve illustrates the form and quality of system performance over the course of a particular disruption and thus shows how well a system has maintained its function with respect to the corresponding initiating disruptive event (Pagano et al. 2019; Förster et al. 2019). The severity of the event outcome and, in return, how well a system has handled the event, can then be quantified based on target functions or summary metrics which capture specific characteristics of the curve, for example, the loss of performance compared to the pre-disturbance performance level (indicating the extent of absorption, Shinozuka et al. 2004), the slope of performance increase (indicating the speed of recovery, Nan and Sansavini 2017), or the area under the curve over the entire disruption (indicating the combined influence of absorption and recovery, Bruneau et al. 2003) – for extensive reviews on different metrics suggested, see Poulin and Kane (2021), Trucco and Petrenj (2023) or Yodo and Wang (2016a).

A performance curve displays the observable effect or outcome of the interplay between a disruptive event and the process which a system carried out to meet this event (i.e., the resilience process). Accordingly, an analysis of the performance curve allows drawing conclusions about the quality of this process (Figure 2). Conclusions from a single curve are event- and situation-specific since a resilience process (as a particular series of actions) is always adapted to the characteristics of the disruption it has to deal with. This also implies that certain aspects of a system's resilience capacities may not play a role in a particular realization of a disruption – a single process may only cover part of the total resilience capacities. For example, in a flood event where dikes do not flood or break, the sophisticated pumping systems behind the dike might not play a role – but they are still an important asset for the system's overall ability to cope with flooding events. Accordingly, the consideration of the outcome of a single event and the associated performance curve allows to draw inference about the quality of a resilience process involved, but is of limited use for addressing a system's resilience capacities (unless you are interested in those specific capacities which allow the system to deal with this particular event in this particular situation).

While many outcome-based approaches base their assessment on a single curve, there are also works which consider ensembles of performance curves. The motivation for considering multiple curves can be to evaluate a system's response to different types (Ouyang et al. 2012; Schoen et al. 2015), magnitudes (Meng et al. 2018; Mugume et al. 2015) and/or manifestations (Kilanitis and Sextos 2019; Kong et al. 2023; Badr et al. 2023) of disturbances, to different failure modes (Diao et al. 2016), or in order to consider the potential variability in a system's response to a particular threat (Galbusera et al. 2018; Meng et al. 2018; Mugume et al. 2015). By considering multiple curves and thus multiple resilience processes, corresponding approaches are better suited to cover a greater share of aspects that make up a system's general capacity to cope with disruptive events, i.e., they can actually provide insights into a system's resilience capacities which underly the possibly diverse manifestations of resilience processes.

## 3.2 The property-based approach

Property-based approaches estimate the status of resilience by assessing the state of system properties or attributes which are assumed to build a system's potential for dealing with disruptive events (see, e.g., Asadzadeh et al. (2017) or Cutter (2016)), i.e., corresponding approaches do not rely on the occurrence of a disruptive event but can be applied to derive information on this potential when the system is not affected by any disruptive event (Rehak et al. 2019). Importantly, this implies that corresponding assessments can be conducted without considering specific scenarios, e.g., by following an 'all-hazards approach' (focusing on a system's core capabilities necessary for dealing with any significant incidence, Petit et al. 2013; Rathnayaka et al. 2024). Although there are also many approaches which target more or less strongly specified disruptive events (Saja et al. 2019; Güngör and Elburz 2024).

Most common within the diverse family of corresponding approaches (see, e.g., Cantelmi et al. (2021), Guo et al. (2021), Bakkensen et al. (2017) or Asadzadeh et al. (2017) for some extensive reviews) are resilience composite indicator frameworks which integrate measurable information from various sources (e.g., survey data, demographic data, expert knowledge, census data) and aggregate them, often following a specific conceptual hierarchy (see, e.g., Storesund et al. (2018), Jovanović et al. (2020), Petit et al. (2013) or Cariolet et al. (2019)), to obtain indices displaying the different aspects of a system's resilience based on different variables (Rehak et al. 2019), themes (Jhan et al. 2020), system functions (Fox-Lent et al. 2015), dimensions (Almutairi et al. 2020) and/or phases of the resilience cycle (Jovanović et al. 2020).

Under the assumption that the (essential) resilience-building system properties are correctly identified, a property-based approach allows to infer conclusions regarding the current status of a system's resilience capacities (Figure 2). For the right assembling of indicators, a deep understanding of the system dynamics under stress is of uttermost importance (i.e., the inclusion of every indicator needs to be properly justified, Cutter et al. 2014). Accordingly, while the construction of a composite indicator might not explicitly rely on the occurrence of a disruption, it still implicitly depends on the hypothetical scenarios which experts have in mind when assessing the importance of alleged resilience properties. Another difficulty with composite (resilience) indicators is how they are combined or aggregated in order to receive a single index or a small set of indices (Greco et al. 2019; Becker et al. 2017; OECD 2008). To provide a valid estimate of the real status of a system's capacity, the calculation of an index had to consider the "correct" weighting of different properties (Lešťáková et al. 2024; Rathnayaka et al. 2024) and the dynamic interplay between them (Gallopín 1996; Cutter et al. 2014; Saja et al. 2019). Since this information is usually difficult to obtain, property-based assessments often represent a useful and applicable way for estimating resilience capacities, which should, however, not be confused with a precise measurement.

# 4. What is preached? The resilience understandings

We propose that there are three main resilience understandings in the literature, i.e., the view of resilience as a process, an outcome, or a capacity. The resilience understandings are reflected in the wording of the presented resilience definition. Resilience is understood as a process when described as "the act of …" (Pimm et al. 2019; Jackson and Ferris 2013; Kanno et al. 2019) or "a process …" (Norris et al. 2008; Scholl and Patin 2014; Ungar 2018). Resilience is understood as an outcome when described as "the speed at which the system returns to …" (Tilman and Downing 1994), "the degree to which …" (Alberti et al. 2003; Diao et al. 2016; Pagano et al. 2019), "how fast …" (Pimm 1984), or "the proportion of affected performance of the system after disruption" (Barker et al. 2013; Shafieezadeh and Ivey Burden 2014; Kong and Simonovic 2019). Resilience is seen as a capacity when described as "the ability to …" (Carlson et al. 2012; Bruneau et al. 2003; NIAC 2010; Rose 2007; Ouyang and Wang 2015; Béné et al. 2012), "the capability to …" (Cai et al. 2018), or "the capacity to …" (Walker et al. 2004; Fiksel 2006). The understanding of resilience as a capacity is the most popular in the context of infrastructure research: the large majority of the 302 definitions of resilience in the context of engineering systems compiled in Wied and colleagues (2020), Mottahedi and colleagues (2021), and Biringer and colleagues (2013) describe resilience as a capacity in their definition (see supplement Table S1 for details).

Depending on how resilience is understood, different possibilities arise for resilience assessment and management (Table 1). In the following, we discuss whether the different understandings allow an assessment solely based on performance data or system properties, enable a straightforward quantification, whether they allow for assessments throughout the full life cycle of infrastructures and whether their assessment is tied to the properties of the disruptive event.

**Table 1: Characteristics of the three main resilience understandings.** The resilience understandings differ with respect to: whether they can be assessed using solely performance data or system properties; allow for straightforward quantification; are continuously defined, i.e., can be assessed at all times, during and in-between disruptions; and are agnostic to the disruptive event. Checkmarks indicate that this criterion is fulfilled by the resilience understanding, minus signs indicate that the criterion is not fulfilled. The checkmark in parentheses indicates that the criterion is only partially fulfilled, for details see discussion sections 5.3.1 and 5.3.2.

| Resilience understanding | Assessable solely via performance data | Assessable solely via system properties | Straightforward quantification | Continuously defined | Agnostic to the disruptive event |
|---|---|---|---|---|---|
| Resilience as a process | ✓ | - | ✓ | - | - |
| Resilience as an outcome | ✓ | - | ✓ | - | - |
| Resilience as a capacity | (✓) | (✓) | - | ✓ | ✓ |

## 4.1 Resilience as a process

In this understanding, resilience is seen as a series of actions that is carried out to meet a disruptive event. An advantage of this understanding of resilience is that, once the corresponding disruptive event has occurred (Keating et al. 2017), this process can be assessed using performance data (Table 1), e.g., conclusions about the quality of the process can be derived from a performance curve (see Figure 2). In addition, the process of handling a specific event can be examined in great detail retrospectively, if the corresponding case study has been documented in sufficient detail (e.g., in a failure report). Thus, the necessary data basis for analyzing the resilience process based on a single event is often relatively readily available. Furthermore, by focusing on a specific case study, it is comparatively straightforward to derive concrete (management) measures that can lead to an improvement of the associated resilience process.

However, the focus on one particular event also implies that the analysis of resilience and corresponding management measures are tied to this one event (the resilience process depends on the characteristics of the event, see section 3.1) and the respective conclusions cannot necessarily be generalized to other events. Accordingly, in this understanding, resilience is unknown as long as the corresponding event is unknown. Therefore, in the understanding of resilience as a process, resilience management is a largely retrospective task which addresses risks that have already been identified. Another implication of the entanglement between resilience process and disruption is that, in this understanding, resilience is not defined in-between disruptions (Ouyang and Wang 2015). The resilience process therefore cannot be assessed in a state of normal operation (Table 1). Furthermore, the resilience process cannot be assessed even during disruptions, as a process has a temporal extent: the series of actions that make up a process must be performed and only when they are completed, the process as a whole can be assessed (although particular sub-aspects of the whole process can be assessed at intermediate points in time during the disruption, see Figure 3). This means that continuous resilience monitoring throughout the full life cycle of a system and therefore operational (acute and demand-based) resilience management, which depends on such monitoring, is not possible (see also section 4.3).

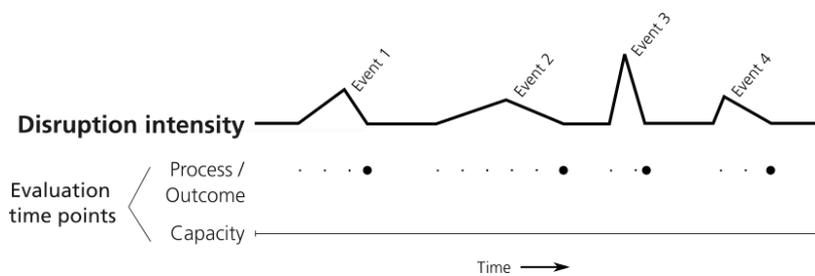

**Figure 3: Evaluation time points for the three resilience understandings.** Depending on the resilience understanding, the time points at which resilience can be assessed differ: When resilience is seen as a process or outcome, it cannot be assessed in-between disruptions, as it is tied to the specifics of the disruptive event. In contrast, when resilience is seen as a capacity, it can be assessed throughout the full lifetime of infrastructures, allowing for continuous resilience monitoring. The filled circles mark the end time points of a disruption where resilience processes and outcomes of that disruption can be assessed as a whole. During the disruption, intermediate outcomes or individual, already completed, sub-processes can potentially be assessed (e.g., absorption), as indicated by the dots.

## 4.2 Resilience as an outcome

The understanding of resilience as an outcome refers to either a content-related outcome or a measure of that outcome (e.g., the outcome "reliability", with the measure of that outcome "system performs normal 99% of the time"). Thus, this understanding of resilience is often very concrete, such that suitable indicators are either directly included in the definition or can easily be derived, e.g., the "speed of return to equilibrium" (Pimm 1984; Tilman and Downing 1994), or "the degree to which the system minimizes level of service failure magnitude and duration" (Butler et al. 2014; Butler et al. 2016). A great advantage of this understanding is that the assessment of resilience is closely tied to its definition, i.e., resilience is easy to grasp, measure and interpret (Table 1).

The unambiguousness and tangibility of understanding resilience as an outcome also has advantages for resilience management: The specification of the desirable outcome clearly states the motivation and gives a concrete target value for corresponding management measures (e.g., "the degree to which the system minimizes level of service failure magnitude and duration", Butler et al. 2014; Butler et al. 2016). It should, however, be noted that, when resilience is seen as an outcome, resilience only exists following a disruptive event (or a series of events, Wang et al. 2024) and is specific to this event (series) (Table 1) – similar to the understanding of "resilience as a process" (Figure 3). In fact, the event-specificity is even more pronounced as the outcome includes its inducing disruptive event (i.e., the outcome is the result of the disruptive event and the associated system response, Wang et al. 2024). This means that the severity of the event (series) that has occurred always affects the evaluation of the success of management measures (i.e., an extreme event probably leads to low resilience). This is reasonable only if mitigating and preventing severe events is seen as part of the resilience concept, a premise that is increasingly challenged (Linkov et al. 2014; Walker 2020; Mentges et al. 2023).

### 4.3   Resilience as a capacity

When resilience is seen as a capacity or a set of capacities, the major drawback is that it is not straightforward to quantify it (since a system's capacities cannot be assessed directly, see Figure 2), and it is only to a limited degree assessable via performance data (Table 1). However, several options remain to assess the resilience capacities indirectly, e.g., by drawing inference about them based on their manifestations (resilience processes, see section 5.3.1), their driving factors (resilience properties, see section 5.3.2), or a combination of the two (see section 5.3.3). In this regard, understanding resilience as a capacity allows to draw from the full spectrum of available assessment approaches.

This understanding of resilience has two major practical advantages. First, resilience is seen as a set of readily available skills, that exist (and are in theory assessable) at any time, i.e., during normal operation as well as during disruptions. Thus, when this understanding of resilience is applied, resilience is a dynamic and ever-present entity that can be tracked or monitored continuously. By monitoring this entity, it is possible to recognize promptly when management interventions ("corrective actions", Curt and Tacnet 2018) are required and then implement them accordingly. Second in this understanding, resilience is not disruption-specific (Table 1). The resilience capacities enable the system to deal with a broad variety of disruptive events, and do not require the identification of high-probability or high-impact events beforehand. Accordingly, this understanding facilitates the notion that resilience might concern unexpected, unknown or, simply, any events (Linkov et al. 2014; Petersen et al. 2020).

## 5. Discussion: Our view on resilience

We argue here that resilience should best be seen as a capacity – a perspective that is also expressed in the majority of scientific publications (see compilations in Wied and colleagues (2020), Mottahedi and colleagues (2021), and Biringer and colleagues (2013), supplement Table S1 for details) and policy documents (e.g., European Union 2022; UNISDR 2015; DHS 2013; BMI 2024). However, in spite of this widely-accepted view on resilience, the applied assessment of resilience often targets not the resilience capacities, but actually the resilience processes.

### 5.1   Why we think that resilience should be a capacity

We believe that how we understand resilience in the context of critical infrastructures should ultimately help in achieving the goal for which it was originally intended, i.e., it should help enable critical infrastructures to cope with any adverse event. In this respect, we think that resilience is best understood as a capacity (Levine 2014). Importantly, this also implies that we see resilience neither as a process nor as an outcome. In the following, we give three reasons why.

REASON #1: RESILIENCE IS NOT RISK.
A central motif in many resilience-related studies is the realization that severe disruptive events can be unforeseen as well as unavoidable and that it is therefore necessary to be prepared to cope with them once they occur (Bi et al. 2023; Wied et al. 2021). It is this idea of being able to deal with anything that distinguishes resilience management from risk management. This means that resilience should be agnostic to any (specific) type of disruption (Mentges et al. 2023) and should in particular allow to address unexpected and unknown events (Linkov et al. 2014; Petersen et al. 2020). If we see resilience as a process or outcome, it is closely tied to the disruption and directly reflects properties of the disruption. If we see it as a capacity, it is less dependent on the disruption but refers to a system-immanent entity. Accordingly, this view highlights that the evaluation of a system's resilience should not be dependent on what has actually happened to the system.

### REASON #2: RESILIENCE IS DYNAMIC.

While resilience is static when viewed as an outcome (a fixed measure determined at the end of a disruption) and has a finite temporal extent when viewed as a process (a series of actions conducted during a disruption), the view of resilience as a capacity highlights that resilience is an ever-present entity that changes and evolves over time (Béné et al. 2012; Wang et al. 2024). This has practical implications. First of all, when resilience is continuously defined, it can theoretically be assessed at any time, i.e., both during and in-between disruptions (Kamissoko et al. 2023; Flanigan et al. 2022). Accordingly, resilience monitoring, defined as "live" tracking of resilience (Kamissoko et al. 2023), is possible. This is an important prerequisite for operational (i.e., "live" and demand-based, Mentges et al. 2023) resilience management as it allows to recognize when management interventions are necessary. Furthermore, understanding resilience as a dynamically evolving capacity (Levine 2014) provides a template for what its long-term or strategic management should look like. A capacity needs to be constantly nurtured and trained in order to be at its peak when it is needed (Moser et al. 2019; Brown and Westaway 2011). This means that managing resilience is not something that can be discontinued at the end of a specific event but a permanent task.

### REASON #3: RESILIENCE IS NOT PERFORMANCE.

The close connection between the performance curve on the one hand and the outcome of the resilience process on the other hand, led to the perception that resilience is often inextricably linked to performance, even to the extent that resilience is sometimes equated with performance (see Box 2). We believe that viewing resilience as a capacity helps to highlight that resilience is somehow linked to performance but, at the same time, helps to distinguish it from performance: it makes clear that resilience is a stand-alone concept that can be assessed even without the use of any performance data (see Figure 2). If we confuse resilience and performance, this has significant implications for resilience management. The selection of measures will strongly differ whether we decide to optimize performance (i.e., processes) or the fundamental ability to deal with unforeseen disruptions (i.e., capacities): in many systems, there is a trade-off between performance and resilience (Homayounfar et al. 2022; Ganin et al. 2017; Pettit et al. 2010). This partly results from the need to weigh the optimization of dealing with one specific disruption against the ability to deal with a large variety of different disruptions (Janssen and Anderies 2007). We claim that understanding (and assessing) resilience as a process or outcome fosters the disproportionate focus on the impacts of single events. Therefore, we advocate to understand resilience as a capacity.

**Box 2: The difference between resilience and performance dynamics.**

The assumption that resilience and performance dynamics "are basically the same" is quite prevalent in the literature. For example, this can be seen when resilience is equated with properties of the performance curve, e.g., the area below the performance curve (Shafieezadeh and Ivey Burden 2014) (the authors explicitly state "resilience is tied with system performance within the period of interest"), or the performance level at the end of the disruption (Cimellaro et al. 2009). In other works, the line between resilience and performance is blurred, for example through terming the performance curve "resilience curve" (e.g., (Flanigan et al. 2022; Poulin and Kane 2021; Gasser et al. 2019; Carrington et al. 2021; Panteli and Mancarella 2015; Oboudi et al.

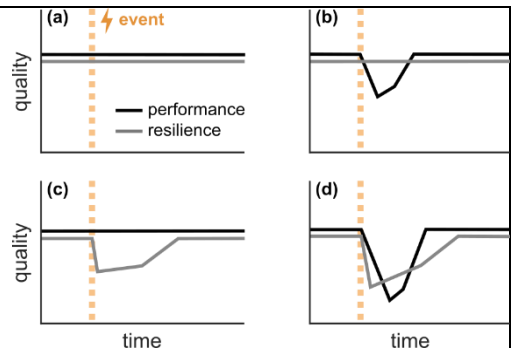

**Figure 4: Difference between performance and resilience dynamics.** Four possible outcomes of an adverse event (dotted orange line) with regard to system performance (black line) and resilience (grey line, where resilience is 2019; Paul et al. 2024) or by tying the dynamics of resilience to the dynamics of performance (Ouyang et al. 2012; Ouyang and Wang 2015; Yodo and Wang 2016b; Fang et al. 2016).

We suggest that resilience and performance dynamics are not necessarily tied together, but each shows its individual signature over time (Figure 4) – although the resilience status evidently has an impact on the maintenance of performance. Few studies explicitly consider the dynamics of resilience and show them alongside the performance curve (Rehak et al. 2019). Moreover, if they do, they usually assume that performance and resilience behave in a similar way; or they consider several dynamic aspects of resilience for quantification purposes, but do not consider how those resilience aspects affect the system's potential to deal with disruptions (i.e., for example, neglecting the yellow arrows "restoration" and "adaptation" in Figure 1b; see Panteli et al. 2017; Gerges et al. 2023; Wang et al. 2024). Accordingly, the possibility of events that primarily or exclusively reduce resilience is largely neglected in the existing literature. In practice, this could have serious consequences, especially if a first disruptive event that primarily affects resilience is followed by further disruptions (i.e., compound events). In such cases, system operators might miss the increased susceptibility of the system. Therefore, we think it is important to be aware and acknowledge that resilience and performance dynamics do not necessarily correlate, but can show contrasting dynamics.

understood as a capacity): **(a)** no impact, system performance and resilience are insensitive to the event (i.e., complete resistance, see Rehak et al. 2018); **(b)** only system performance is affected by the event (e.g., performance of a wind farm drops due to lack of wind, resilience remains unchanged by this change in weather conditions); **(c)** only resilience is affected by the event (e.g., a storm destroys a redundant power line, resulting in a decrease in the absorptive resilience capacity, while there is no decrease in performance as only the redundant line is affected, see definition of "redundancy" in Haimes 2009); or **(d)** impact on system performance and resilience (e.g., a storm destroys numerous electricity posts, resulting in both decreased system performance, as some parts of the network loose connectivity, and decreased absorptive capacity, as some redundant power lines are no longer functional).

## 5.2 Do we practice what we preach?

Several high-impact works define resilience as a capacity, but assess the resilience process. We do think that assessing and describing the resilience process, as has been done in the following examples, has great merit. However, we also believe that it is important to explicitly discuss the implications of the deviation between the definition ("what is preached") and the assessment approach applied in the respective study ("what is practiced"). For example, Bruneau and colleagues (2003) define resilience as a capacity ("the ability of social units to mitigate hazards, contain the effects of disasters when they occur and carry out recovery activities"), but measure resilience using a single performance curve that shows the outcome of the interplay between one concrete disruptive event and the respective realized resilience process (see section 3.1). Thus, the authors state that resilience is a capacity, but actually measure the quality of a single event-specific resilience process. Accordingly, in order for their measure of resilience to be consistent with their definition of resilience, it must be implicitly assumed that this one process is indicative of the entire spectrum of abilities/activities (and their intrinsic variability) which allow the system to cope with "hazards" and/or "disasters". This implicit assumption, however, is not noted or discussed in the work. Other examples include the works by Ouyang and colleagues (2015), Reed and colleagues (2009), and Shafieezadeh and colleagues (2014), in which resilience is initially defined as an ability (e.g., "ability to resist (prevent and withstand) any possible hazards, absorb the initial damage, and recover to normal operation", Ouyang and Wang 2015), but the measurement of resilience then relies on the evaluation of a single performance curve. Thus, the understanding is that resilience is a capacity, but the assessment assumes it is a process, with the same restrictions applying as for the previous example.

We find that the majority of studies, which fall into the class of outcome-based resilience assessment approaches (see section 3.1), define resilience as a capacity, while the methods which are actually applied often address the resilience process. Thus, in many papers, we see a mismatch between the conceptual understanding of resilience and the methodologic approach for assessing resilience (exceptions are those studies that understand resilience as an outcome, in those studies the definition and assessment of resilience naturally match well, e.g., Diao et al. 2016). The valuable work in these studies could benefit from an additional discussion of the implications or possible limitations when extending conclusions based on a single resilience process to the underlying capacities.

## 5.3 Options for assessing resilience capacities

If we acknowledge that resilience is a capacity, it is more difficult to assess it, compared to a resilience process or an outcome. A resilience capacity is an abstract entity that cannot be directly observed. To understand and quantify it, we need to rely on more tangible measures ("proxies for resilience", Doorn 2017), i.e., (1) the resilience-building system properties (a subset of all system properties) and/or (2) the performance of the system.

Using the resilience properties as indicators can be seen as the "top-down" approach: assessing resilience by analyzing its driving factors (Figure 2). Accordingly, the alternative approach can be seen as the "bottom-up" approach: making inferences about resilience by analyzing its effects on an observable system characteristic (e.g., performance) in one or several specific situations (Figure 2, note that this does not comply with the use of the terms top-down and bottom-up in the context of disaster resilience as outlined in Cutter 2016). The last option is to combine the two, to get a maximally holistic picture of resilience.

### 5.3.1 Option A): Outcome-based approach

We believe that to learn something about the resilience capacities, a single performance curve is not enough. Metrics which are derived from a single performance curve can only describe the quality of one particular resilience process, i.e., they summarize the outcome of the activated resilience capacities in response to one specific situation. If resilience was defined as a system's capacity to cope with this particular event, the assessment of the observed resilience process was sufficient to characterize the system's resilience. However, resilience definitions usually mention the capacity to cope with different events, often even including unforeseen and unexpected events (Mentges et al. 2023). Since a system might be differently well equipped to handle different types and manifestations of disruptive events, information on the quality of multiple resilience processes (i.e., multiple performance curves) are needed to learn about its resilience capacities. For example, it is helpful to include information on the system's reaction to several disruptive events which differ with regard to their type (e.g., windstorm or heatwave) and form (e.g., intensity, spatial characteristics, temporal expansion), in terms of several different performance metrics, or several disruptive events of the same kind at multiple points in time (i.e., different states of the system) – also see the metaphor of resilience as a system's stability landscape (Holling 1973; Walker et al. 2004) and quantifiers inspired by it (Menck et al. 2013; Mitra et al. 2015; Dakos and Kéfi 2022; Bien et al. 2023) which very well illustrate how resilience includes the ability to deal with many different disruptive events. This combination of information allows to extend conclusions from a single resilience process towards the underlying resilience capacities of the system. Others have argued before that for a comprehensive picture of resilience, a large variety of events is needed (Braun et al. 2020), especially including those with low occurrence probability but high impact (Diao et al. 2016).

Generally, an exclusively outcome-based approach allows us to assess whether and to what extent a system's resilience status fulfills pre-defined requirements, which are given by the set of probed disruptive events and the choice of evaluation metrics. Thus, the outcome-based approach can be applied to identify areas for intervention (e.g., when a system is equipped to handle a severe but not an extreme storm surge, additional measures to handle particularly intense events are needed). However, to effectively intervene, one needs to understand which system properties or traits can be manipulated in order to strengthen a system's resilience to meet the requirements established in the outcome-based analysis (i.e., at least a rudimentary sensitivity analysis is required). In order to gain this understanding, one needs to enter the realm of property-based approaches, e.g., by selecting and probing different system configurations.

### 5.3.2 Option B): Property-based approach
The often-utilized composite indicator frameworks are particularly suitable when the selection of indicators is based on a deep understanding of the respective system (and the adverse influences it potentially has to deal with), in particular, when there is broad agreement among system experts about which properties are essential (Levine 2014). In practice, it should be noted that this means that the evaluation of the importance of certain indicators will probably be implicitly process-based, i.e., experienced experts will likely have certain situations and processes in mind on the basis of which they conduct respective judgements. In contrast, truly general or standardized indicators are those derived on the basis of general resilience principles or design criteria (e.g., redundancy, diversity, flexibility, resourcefulness, Cariolet et al. 2019; Sharifi and Yamagata 2016; Sharifi 2023). These, however, can only provide valuable and valid insights into a system's resilience when they are deemed important by local stakeholders and system experts (Cariolet et al. 2019; Almutairi et al. 2020; Güngör and Elburz 2024) – which brings us back to the necessity to base any assessment of resilience on a deep understanding of the system whose resilience one wants to assess.

There are two main issues with the purely property-based resilience assessment. The first is that corresponding indices usually do not consider the potentially complex systemic impact of and interplay between different system properties (Yabe et al. 2022; Cariolet et al. 2019; Saja et al. 2019). Instead, the offsetting of different indicators often follows a simplistic scheme that is not motivated by any consideration of system dynamics or behavior, e.g., applying relative and purely data-driven normalization (e.g., min-max scaling, Cutter et al. 2014; Asadzadeh et al. 2017), equal, arbitrary or subjective weighting (e.g., "improvements in any component of a resilience score are of equal importance", Levine 2014), or linear additive aggregation (i.e., no conditional interplay or complex interdependencies between parameters, Rathnayaka et al. 2024; Yabe et al. 2022). While, due to the complexity of the systems under consideration, a sufficiently detailed description of the precise impact of every system property is hard (if not impossible) to obtain, it is questionable whether an aggregation of different indicators makes any sense at all if it does not mirror the importance and complexity of the real interplay between system properties (Sharpe 2004; Greco et al. 2019; Doorn 2017). A second, related and well-documented issue of exclusively property-based approaches is that they tend to lack (sufficient) validation (Cariolet et al. 2019; Shiozaki et al. 2024; Cutter et al. 2014) – the selection of indicators is typically based on "judgement, rather than empirical evidence" (Levine 2014; Wang et al. 2023). This validation can only be obtained by testing the impact of the proposed indicators in actual or simulated crisis scenarios, i.e., it is necessary to complement the property-based approach with a process- or outcome-based approach (Shiozaki et al. 2024).

### 5.3.3 Option C): Combination of the two approaches
We believe that, ideally, to assess the resilience capacity, property- and outcome-based approaches should be combined in order to harness their respective benefits and to balance their specific deficiencies. In general, property-based approaches can be utilized to uncover potential links between resilience properties and capacities, which are necessary for conducting resilience monitoring and management, and process- and outcome-based approaches allow to verify and further specify these links. For instance, if a process-based model of the system of interest is available, this model can be utilized to examine the influence of different system properties on a system's resilience, i.e., various properties can be compared in some form of sensitivity (Linkov et al. 2018), correlation (Meng et al. 2018) or impact (Levine 2014) analysis where the effect variable is derived from a selection of disruptive scenarios. The knowledge about the influence of specific system properties can then be used to select management actions or to set up monitoring schemes that target the most influential resilience properties.

However, a simulation-based evaluation is often not possible since obtaining an accurate dynamical model is not feasible. In this case, a primarily property-based approach relying on a composite of indicators might be the best option. It is, however, important to validate the resilience-building quality of the selected properties via evaluation of actual instances of a severe disruptive event and its outcome (Bakkensen et al. 2017; Keating et al. 2017; Burton 2014). If such data is missing, system experts can be consulted to perform the validation. When interviewing system experts regarding the importance of resilience-building parameters, knowledge of the resilience processes is likely implicitly taken into consideration. However, it would be even better to include this more explicitly. One way to do this is to discuss several storylines, i.e., exemplary hypothetical crisis scenarios (Bruijn et al. 2016; Munz et al. 2023; Brito et al. 2024), with experts, to validate which system properties actually play a role in the process of dealing with the notional situation.

## 6. Conclusion

Here, we identify three main resilience understandings from the literature: the understanding of resilience as a process, an outcome, and a capacity. We argue that each of the understandings has certain advantages and drawbacks. The understanding of resilience as an outcome or process facilitates the assessment of resilience, as suitable indicators can easily be derived. However, only the understanding of resilience as a capacity allows for resilience monitoring, operative resilience management, and strategic resilience management in the stricter sense. Furthermore, only this understanding preserves a key property of resilience: resilience should be indifferent towards any specific disruption which the system has to deal with. We find that while the majority of resilience definitions refer to resilience as a capacity, the prevailing performance-based approaches to assess resilience target either the resilience process or the outcome of this process. Thus, there is a conflict between the proposed conceptual understanding of resilience and the chosen method of evaluation. We argue that the applied assessment approach should match the claimed understanding. The challenge here is to resist "switching" the understanding of resilience between the introduction and the methods section for the benefit of, on the one hand, a very universal and comprehensive definition of resilience but, on the other hand, a practical and feasible assessment method that uses readily available data.

We agree with previous studies that it is useful to analyze performance data and to learn from past disruptions. Although quantifying resilience capacities is a daunting task, it is still worth tackling and making due with what (limited) data we have: usually, one or few performance curves. However, we emphasize that when resilience assessments do not adequately reflect the conceptual understanding of resilience, this needs to be noted and the implications and limitations of interpretation need to be discussed. To avoid any doubts, it should always be stated what aspect of the resilience framework (e.g., which component of the "resilience backbone") is considered and how this affects the scope of validity of the respective study findings. Furthermore, we believe that the consideration of a single or a few performance curves is of little use if it is not accompanied by an analysis of why the system reacted the way it did. Obtaining such an understanding requires looking into the dynamics of the system's response (the actual resilience process) as well as into the specifics of the disruptive event that caused this response (including its severity). If the latter is left unconsidered, this could mean a system's resilience is graded according to its recent level of exposure. Thus, instead of preparing for "anything that could possibly happen", we might end up being prepared for "what just happened". In this respect, we believe that consequently treating resilience as a capacity will help prepare for all kinds of surprises coming our way.

**Acknowledgements**
We thank Walaa Bashary for helpful feedback on an earlier version of the manuscript.

# 7. Publication bibliography

## Supplementary information

**Title of article**: Do we practice what we preach? The dissonance between resilience understanding and measurement

**Authors**: Lukas Halekotte, Andrea Mentges, Daniel Lichte

### S1: Relation to the distinctions presented in Moser et al. 2019

In their work on the "turbulent world of resilience", Moser et al. (2019) distinguish between three main interpretations of resilience in the literature – resilience as a system trait, a process or an outcome – based on the focus of different resilience-related studies, i.e., whether they focus on characteristics which make a system resilient (trait/property), on the actions a system needs to carry out to behave resilient (process) or on the desirable outcome(s)/states a system should achieve (outcome). The first group comprises works that focus on resilience as a system trait, with the goal of identifying system characteristics (e.g. redundancy, rapidity, adaptive capacity). This group would correspond to a combination of understanding resilience as a capacity or resilience property. In contrast to our theory, which says that resilience is a capacity which is built by certain system properties, their grouping does not make this distinction. The second category comprises works that focus on actions and interventions, i.e. resilience management, however e.g. "embrace change", "integrate local knowledge". This corresponds in our terminology to the resilience processes and adaption processes. In contrast to our understanding of "resilience as a process", they do not focus on the process of dealing with a disruption but the process of adapting/re-structuring that need to be performed continuously to work towards some goal. The third category, "outcome" is described as works that are focused on "(temporary) states of a system", i.e. monitoring (e.g. reliability, decreased vulnerability). Outcome in our case refers to the use of a quantifiable measure for the effects of a disruptive event; in their case it refers to a conceptual / abstract goal, i.e. the result of the resilience-enhancing efforts. Thus, our understanding of resilience as a capacity or ability incorporates all of these approaches since each aspect (property, process, outcome) has a direct or indirect link to the resilience capacity (see causal relationships in the resilience backbone and Figure 1).

Their distinction is mainly derived as part of a larger effort to bring structure into the myriad of resilience papers; it is meant as a grouping mainly in the sense of clustering works with a similar focus together or distinguishing them. The practical concrete consequences for assessment, monitoring and management of choosing either one of the groupings are not the focus of their work.

### S2: How are the resilience understandings represented in the literature?

To tackle this question, reviewed a total of 302 definitions of resilience, compiled in works by Wied and colleagues (2020), Mottahedi and colleagues (2021), and Biringer and colleagues (2013). We classified the definitions based on the wording and recorded whether they describe resilience as a capacity (i.e. resilience is "the ability to..", "the capacity to.."), a process ("the action of", "acting .."), a quantifiable measure (, "the speed of..", "the rate of..", "the number of.."), or a mixture of these (see full table Table S1).

We found 21 statements which define a resilient system rather than resilience itself (Table S1). We classified these statements as "indirect", because a process may be used to indicate that a system is resilient, while resilience itself is understood as a capacity. This distinction would not be apparent in a statement on how to identify resilient systems. Therefore, such statements cannot be used as a basis to answer our question. Another 7 statements were too short or too vague to allow a clear assignment to one of our classification categories. We marked these statements as "unclear". Also, we emphasize that some of our classifications may not reflect the acutal view of the associated paper, as the classification was based on the short excerpts chosen by the authors of the compilation studies. An example is the statement from Usdin (2014). In the supplementary material of Wied and colleagues, the definition of resilience by Usdin is summarized as "how systems learn from and adapt to trauma-related experiences", whereas in the original publication, Usdin precedes this statement with a half-sentence which emphasizes the resilience capacities, i.e. "the definition has been modified over time to include the adaptive or learning qualities of the system – how systems learn from and adapt to trauma-related experiences". Thus, conclusions drawn based on the collection of short statements in the studied reviews may be flawed due to the restricted context which was assessed. Nonetheless, we think that a general tendency of the understanding is likely reflected in the excerpts. Also, the overall numbers need to be treated carefully as definitions can occur multiple times when they are included in more than one of the studied reviews.

Overall, we found that in all three reviews, most definitions describe resilience as a capacity (71% in Wied and colleagues, 85% in Mottahedi and colleagues, and 57% in Biringer and colleagues, Table S), while only few consider resilience a process (4%, 0%, and 9%, respectively) or a measure (9%, 7%, and 22%, respectively). Note that this distribution differs strongly in other disciplines, e.g. in psychology the view of resilience as a process is prevailing (Ungar 2018).

**Table S1: Scope of resilience definitions.** We reviewed the definitions of resilience compiled in Wied et al. (2020, n=251), Mottahedi et al. (2021, n=28), and Biringer et al. (2013, n=23). The definitions either view resilience as a capacity, a measure, a process, or a combination. Some of the definitions did not define resilience itself, but characterized a resilient system (indirect). Other definition text excerpts were too short to allow a classification (unclear).

| Scope of definition | Wied et al. (2019) Occurr. | Percent. [%] | Mottahedi et al. (2021) Occurr. | Percent. [%] | Biringer et al. (2013) Occurr. | Percent. [%] |
| --- | --- | --- | --- | --- | --- | --- |
| Capacity | 179 | 71.1 | 24 | 85.7 | 13 | 56.6 |
| Measure | 22 | 8.8 | 2 | 7.1 | 5 | 21.7 |
| Indirect | 21 | 8.4 | 2 | 7.1 | 2 | 8.7 |
| Process | 9 | 3.6 | 0 | 0 | 2 | 8.7 |
| Capacity + Process | 9 | 3.6 | 0 | 0 | 1 | 4.4 |
| Unclear | 7 | 2.8 | 0 | 0 | 0 | 0 |
| Capacity + Measure | 4 | 1.6 | 0 | 0 | 0 | 0 |

# 1. Publication bibliography